\documentclass[12pt]{article}
\usepackage{amssymb,amsmath}
\usepackage{cite}
\newlength{\dinwidth}
\newlength{\dinmargin}
\newtheorem{theorem}{Theorem}[section]
\newtheorem{prop}[theorem]{Proposition}
\newtheorem{lemma}[theorem]{Lemma}
\newtheorem{cor}[theorem]{Corollary}
\newtheorem{definition}[theorem]{Definition}
 
\newenvironment{proof}{\medskip \noindent 
            {\bf Proof.}}{ \hfill $\square$ \medskip}

\newcommand{\ie}{{\it i.e.\ }}

\newcommand{\bnull}{{\mbox{\boldmath $0$}}}
\newcommand{\bb}{{\mbox{\boldmath $b$}}}
\newcommand{\be}{{\mbox{\boldmath $e$}}}

\newcommand{\br}{{\mbox{\boldmath $r$}}}
\newcommand{\bx}{{\mbox{\boldmath $x$}}}

\newcommand{\sbe}{{\mbox{\footnotesize \boldmath $e$}}}

\newcommand{\blambda}{{\mbox{\boldmath $\Lambda$}}}

\newcommand{\Lp}{{\cal L}_+}
\newcommand{\Lpo}{{\cal L}_+^\uparrow}

\def\idty{{\leavevmode\hbox{\rm 1\kern -.3em I}}}

\def\wrnet{{\{\Rs(W)\}_{W \in\Ws}}}

\def\Cs{{\cal C}}

\def\Hs{{\cal H}}

\def\Js{{\cal J}}

\def\Ls{{\cal L}}

\def\Ps{{\cal P}}

\def\Rs{{\cal R}}

\def\Ts{{\cal T}}

\def\Ws{{\cal W}}

\def\Pid{{\Ps_+ ^{\uparrow}}}
\def\Lid{{\Ls_+ ^{\uparrow}}}

\def\idty{{\leavevmode\hbox{\rm 1\kern -.3em I}}}

\def\RR{{\mathbb R}}
\def\CC{{\mathbb C}}

\def\IN{{\mathbb N}}
\def\ZZ{{\mathbb Z}}
\def\bx{{\mbox{\boldmath{$x$}}}}

\def\beq{\begin{equation}}
\def\eeq{\end{equation}}
\begin{document}
\title{An Algebraic Characterization of Vacuum States \\
in Minkowski Space.\ III.\ Reflection Maps}
\author{{\Large Detlev Buchholz\,$^a$ \ and \  
Stephen J.\ Summers\,$^b$ }\\[5mm]
${}^a$ Institut f\"ur Theoretische Physik, 
Universit\"at G\"ottingen, \\ 37077 G\"ottingen, Germany  \\[2mm]
${}^b$ Department of Mathematics, 
University of Florida, \\ Gainesville FL 32611, USA}

\date{} 

\maketitle 

{\abstract \noindent Employing the algebraic framework
of local quantum physics,\! vacuum states in Minkowski space are
distinguished by a property of geometric modular action. 
This property allows one to construct from any locally generated
net of observables and corresponding state a continuous unitary
representation of the proper Poincar\'e group which acts covariantly
on the net and leaves the state invariant. The present results and
methods substantially improve upon previous work. In particular, the
continuity properties of the representation are shown to be a
consequence of the net structure, and surmised cohomological
problems in the construction of the representation are resolved 
by demonstrating that, for the Poincar\'e group, continuous 
reflection maps are restrictions of continuous homomorphisms.}

\section{Introduction}

\setcounter{equation}{0}

A basic conceptual problem in local quantum physics \cite{Haag} is
the determination of the spacetime symmetry, causality and stability 
properties of a theory from the structure of the observable algebras
associated with spacetime regions. Within this general 
setting, it seems unnatural to appeal from the outset to 
symmetry properties of a theory --- such as the action of a spacetime isometry
group upon the states and observables --- which are absent in generic 
spacetimes. Instead, the pertinent notions characterizing specific 
physical systems ought to be based on the states and observables
of the system, and possible symmetry and stability properties 
should be deduced, not posited.

Light was shed on these matters by a condition of
geometric modular action (CGMA), proposed in \cite{BS,BDFS} and
briefly recalled in Section 4, which is designed to characterize those
elements in the state space of a quantum system which
admit an interpretation as a ``vacuum''. This condition is
expressed in terms of the modular conjugations associated to any given
family of algebras paired with suitable subregions (wedges) of the
underlying space--time and any states by the Tomita--Takesaki
modular theory, cf.\ \cite{BratRob,KadRing}. It thereby can be applied,
in principle, to theories on any space--time manifold.  For a
motivation of this condition and applications to theories in
Minkowski, de--Sitter, anti-de Sitter and a class of Robertson--Walker
space--times, we refer the interested reader to
\cite{BDFS,BFS,BMS1,BFS1}. 

In the present article we revisit the case of Minkowski space
theories and resolve some intriguing questions left open by
our previous work in \cite{BDFS,BFS}. The basic 
ingredients in that investigation are an isotonous map 
(henceforth, a net) 
$W \mapsto \Rs(W)$ from the family of wedge-shaped regions $W \subset \RR^4$,
bounded by two characteristic planes, 
to von Neumann algebras $\Rs(W)$
on a Hilbert space $\Hs$, and a state vector $\Omega \in \Hs$ 
complying with the CGMA. The modular conjugation 
associated to any given pair $\big(\Rs(W), \Omega\big)$ 
was shown to have the geometrical meaning 
of a reflection $\lambda$ about the edge of $W$. More precisely,
denoting this conjugation by $J(\lambda)$, one has for any 
wedge $W_0$ the equation \cite{BDFS}
\begin{equation} \label{covar}
J(\lambda)  \Rs(W_0) J(\lambda) = \Rs(\lambda W_0) \, .
\end{equation}
This implies, in particular, that $J(\lambda)$ is also the 
modular conjugation of the pair $\big(\Rs(W^\prime), \Omega\big)$,
where $W^\prime = \lambda W$ denotes the causal complement of $W$;
so the net satisfies wedge duality. Moreover, one has 
\begin{equation} \label{prop}
J(\lambda)  J(\lambda_0)  J(\lambda)  =  J(\lambda\lambda_0\lambda) \, , 
\end{equation} 
in an obvious notation. As the reflections $\lambda$ generate the 
proper Poincar\'e group $\Ps_+$, these relations 
lead naturally to the question of whether products of
the conjugations $J(\lambda)$ generate an (anti) 
unitary representation of  $\Ps_+$ which acts covariantly on the net. 

This question was answered in the affirmative in \cite{BDFS,F2} 
under a technical assumption of net continuity. 
It follows from that assumption that the map
\begin{equation}  \label{cont}
\lambda \mapsto J(\lambda) \end{equation}
from the reflections in 
$\Ps_+$ into the group of (anti)unitary operators is continuous.
This information, together with relation (\ref{prop}),  
implies that there is a continuous
projective representation of  $\Ps_+$ on $\Hs$
with coefficients in the center of the 
group $\Js$ generated by all conjugations $ J(\lambda)$. By an
application of Moore cohomology theory, this projective representation 
lifted to a true representation and the center of $\Js$ 
turned out to be trivial \cite{BDFS}.

These latter results suggested that viewing the
problem cohomologically was misleading and obscured the presence of an
extremely rigid structure encoded in the modular conjugations. It was
also desirable to clarify the conceptual status of the technical
assumptions underlying the crucial continuity property of the map
(\ref{cont}). A first step towards the clarification of these points
was taken in \cite{BFS}. There it was shown without any {\it a priori}
continuity assumptions that a continuous
unitary representation of the subgroup of translations acting
covariantly on the net can be constructed from the modular
conjugations. In the present investigation we want to 
extend this result to the full proper Poincar\'e group
$\Ps_+$. 

We shall restrict our attention here to nets $W \mapsto \Rs(W)$ which are
locally generated, a case of particular interest being the situation
where each $\Rs(W)$ is the inductive limit of algebras $\Rs(C)$
associated to double cones $C \subset W$.  In fact, this condition is
already satisfied if $\Omega$ is cyclic for the algebras
$\Rs(C)$ and satisfies, in addition to the CGMA, a modular stability
condition (CMS), recalled in Section 4, which was 
proposed in \cite{BDFS} for the characterization of stable states.
We shall show under these latter two
conditions that the map (\ref{cont}) provided by the
CGMA is continuous.

Our second result clarifies the nature of continuous maps  
(\ref{cont}) of reflections $\lambda \in \Ps_+$ into 
the topological group $\Js$, which satisfy the basic relations
$J(\lambda)^2 = 1$ and
$J(\lambda)  J(\lambda_0)  J(\lambda)  =  J(\lambda\lambda_0\lambda)$
for any pair of reflections $\lambda, \lambda_0 \in \Ps_+$.
We shall show that such reflection maps  
are restrictions of continuous homomorphisms from 
$\Ps_+$ into  $\Js$. Phrased differently, 
any reflection map can be extended uniquely 
to a true continuous (anti)unitary 
representation $U$ of $\Ps_+$ which, in view of (\ref{covar}), acts
covariantly upon the net.

Thus the outcome of the present investigation is the insight that
any vector $\Omega$ which is cyclic for the local algebras and
complies with the CGMA and the CMS is a vacuum state which is
invariant under a continuous (anti)unitary representation of $\Ps_+$
acting covariantly on the net. No further assumptions are
needed for the proof of this result. That this arises in the manner
shown here provides further evidence that the
modular involutions are fundamental objects encoding crucial physical
data, which include the causal structure of the theory, its dynamics,
and the action of the isometry group upon the observables.  Even the
space--time itself can be found to be encoded in the modular
involutions in certain cases \cite{SW}.

Our paper is organized as follows. In the next section we consider
continuous reflection maps from the proper Lorentz group into
an arbitrary topological group and show that they are
restrictions of continuous homomorphisms. The continuity of the
reflection maps arising in the present context is established in
Section 3.  These results are combined in Section 4 with theorems we
have previously established to yield the desired
characterization of Poincar\'e covariant vacuum states in Minkowski
space in terms of the modular objects.

\section{Reflection maps and homomorphisms}

\setcounter{equation}{0}

In this section  we study continuous reflection maps from 
the proper Lorentz group into an arbitrary topological 
group. We shall show that any such map is the res\-triction
of a continuous homomorphism. Combining this information 
with results obtained in \cite{BFS}, this feature can  
be established also for reflection maps on the proper
Poincar\'e group. In fact, similar results hold 
for many other groups, suggesting the possibility of a general 
theorem about reflection maps. This would be of  
interest in the general context of the CGMA, where reflection
maps appear naturally \cite{BDFS}, and in the application of the CGMA
to other space--times. But we shall not address 
the general problem here.

\subsection{Group theoretical considerations}

     Let $\Lp$ be the proper Lorentz group and $\Lpo$ be its
orthochronous subgroup. Fixing a Lorentz system with proper
coordinates  $(x_0,\bx) \in \RR^4$ and metric in diagonal form 
$g = \mbox{diag} (1,-1,-1,-1)$, one can uniquely decompose any
$\Lambda \in \Lpo$ into a rotation $R$ in the time-zero plane and
a boost (velocity transformation) $B$,
\begin{equation}
\Lambda = R B.
\end{equation}
{\bf Remark:} This formula is simply the polar decomposition of
$\Lambda$ in the space $M(4,\RR)$ of real four--by--four matrices. 
In particular, any Lorentz transformation which is represented by a positive
matrix is a boost. This well--known fact recently received some 
attention again, cf.\ \cite{Mor,Urb}.

     Thus, any $\Lambda \in \Lpo$ generically fixes two spatial
directions, the axis of revolution $\br$ of $R$ and the boost
direction $\bb$ of $B$. We adopt the convention that $\br$, $\bb$ are
normalized and that rotations are performed by an angle less than or
equal to $\pi$ about $\br$ in the counterclockwise direction. So $\br$ is
fixed unless $R=1$ or $R$ is a rotation through the angle $\pi$, where
$\br$ is fixed only up to a sign. Similarly, unless $B=1$,
the direction of $\bb$ is fixed by the condition that 
the lightlike vector $(1,\bb)$ 
is an eigenvector of $B$ corresponding to its eigenvalue
which is larger than 1.

     Making use of this convention, we want to show that any $\Lambda$
can be represented as the product of two reflections about the edges
of suitable wedges. Although there are results in the mathematical
literature which establish that any element of the Lorentz group
can be written as the product of two involutions, the reflections 
we must employ are restricted to lie in 
a single conjugacy class of these involutions. 
We are therefore obliged to provide a proof of this fact here.

     We begin by defining for given unit vector $\be \in \RR^3$ the wedge
\begin{equation}
W_\sbe \doteq \{ x \in \RR^4 \mid \bx \cdot \be > |x_0| \}
\end{equation}
and the involution $\lambda_\sbe \in \Lp$ inducing the 
reflection about its edge, \ie
\begin{equation}
\lambda_\sbe \, (1, \bnull) = - (1, \bnull)\ , \quad 
\lambda_\sbe \, (0, \be) =  - (0, \be) \ , \quad 
\lambda_\sbe \, (0, \be_\perp) =  (0, \be_\perp) \ ,
\end{equation}
where the latter equality holds for any $\be_\perp$ which is perpendicular 
to $\be$. One finds that if $\bb \cdot \be = 0$ and $B$ is any boost in 
the direction of $\bb$, then $\lambda_\sbe \, B = B^{-1} \, \lambda_\sbe$ 
is a reflection about the edge of the boosted wedge $B^{-1/2} \, W_\sbe$. 
Similarly, if $\br \cdot \be = 0$ and $R$ is any rotation about the 
direction of $\br$, then  $R \, \lambda_\sbe =  \lambda_\sbe \,  R^{-1}$
is a reflection about the edge of $R^{1/2} W_\sbe$,
where $R^{1/2}$ is the rotation about $\br$ through half the angle of $R$. 
Thus, choosing for given $R$, $B$ the direction $\be$
such that $\br \cdot \be = \bb \cdot \be = 0$, $R \, \lambda_\sbe$
as well as $\lambda_\sbe \, B$ are reflections about the edges of 
wedges and $R \, \lambda_\sbe \, \lambda_\sbe \, B = R B = \Lambda$.

     In the following we shall call the involutions $\lambda$ inducing
reflections about the edges of wedges simply reflections, for short,
and we shall denote by $\Rs$ the set of all such reflections. We have
therefore just proved the following result which has been
proven independently by Ellers \cite{El} using a very different
argument.
\begin{lemma} Every element of the proper orthochronous Lorentz group
can be written as a product of two reflections, \ie for every 
$\Lambda \in \Lid$ there exist two elements $\lambda_1,\lambda_2 \in \Rs$
such that $\Lambda = \lambda_1 \lambda_2$.
\end{lemma}
     This result is crucial in our investigation of reflection maps
on the Lorentz group, and a similar result is likely to be just as
important in any attempt to generalize our results to other groups.
We refer the interested reader to the recent paper of Ellers \cite{El}
for a beginning of such a program.

     We shall discuss the ambiguities involved in this representation.
Let $\Lambda \in \Lid$ be given and let $\lambda_1, \lambda_2 \in \Rs$ 
be reflections
such that $\lambda_1 \lambda_2 = \Lambda$. If $\Lambda^\prime \in \Lpo$ 
is any Lorentz transformation commuting with $\Lambda$, it is clear
that the product of 
the reflections $\lambda_1^\prime \doteq \Lambda^\prime \lambda_1
\Lambda^{\prime \, -1}$, $\lambda_2^\prime \doteq \Lambda^\prime \lambda_2
\Lambda^{\prime \, -1}$ is equal to $\Lambda$. Yet this may not be the 
only ambiguity in the choice of pairs of reflections corresponding to
$\Lambda$. Since ${\lambda_1}^2 = 1$, one has
$\lambda_2 = \lambda_1 \Lambda$, and since ${\lambda_2}^2 = 1$, one
also has $\lambda_1 \Lambda = \Lambda^{-1} \lambda_1$. Now if
$\lambda_1^\prime$ is another reflection satisfying the 
latter equation, one gets 
$\lambda_1 \lambda_1^\prime \Lambda = \Lambda  \lambda_1 \lambda_1^\prime$,
\ie $\lambda_1^\prime = \lambda_1 \Lambda^\prime$, where $\Lambda^\prime$
commutes with $\Lambda$. Moreover, as $\lambda_1^\prime$ is
an involution, $\Lambda^\prime$ must satisfy 
$\lambda_1 \Lambda^\prime = \Lambda^{\prime \, -1} \lambda_1$. 
We therefore consider for given $\Lambda$ and reflection $\lambda_1$ as 
above the set of Lorentz transformations
\begin{equation}
\blambda^\prime = 
\{ \Lambda^\prime \in \Lpo \mid \lambda_1 \Lambda^\prime 
= \Lambda^{\prime \, -1} \lambda_1, \
\Lambda^\prime \Lambda = \Lambda \Lambda^\prime \} \ .
\end{equation}
Given any $\Lambda^\prime \in \blambda^\prime$, the elements 
$\lambda_1^\prime \doteq \lambda_1 \Lambda^\prime$ and 
$\lambda_2^\prime \doteq \lambda_1 \Lambda^\prime \Lambda$ are involutions, 
and their product is equal to $\Lambda$. But these involutions are  
not always reflections. Nonetheless, if $\Lambda$ is such that  
for each  $\Lambda^\prime \in \blambda^\prime$ there is some 
$\Lambda^{\prime \, 1/2} \in \blambda^\prime$ whose square is $\Lambda^\prime$,
one has 
$\lambda_1^\prime = \Lambda^{\prime \, -1/2}  \lambda_1  
\Lambda^{\prime \, 1/2} $ and 
 $\lambda_2^\prime = \Lambda^{\prime \, -1/2} \Lambda^{-1/2}  \lambda_1 
\Lambda^{1/2} \Lambda^{\prime \, 1/2} $. So, in this case, 
$\lambda_1^\prime$ and $\lambda_2^\prime$ are both reflections 
of the form given above.

     In the following we shall focus our attention on certain specific 
elements $\Lambda \in \Lpo$ which are of the form 
$\Lambda = \Lambda_1 \Lambda_0 \Lambda_1^{-1}$,
where $\Lambda_1 \in \Lpo$ is arbitrary and 
$\Lambda_0$ (${\Lambda_0}^2 \neq 1$) is an element of 
the stability group ${\cal L}_0 \subset \Lpo$ of 
some fixed wedge $W_{\be_0}$. We recall that ${\cal L}_0$
is the abelian subgroup of $\Lpo$ generated by all rotations $R_0$
about $\be_0$ and all boosts $B_0$
in the direction of $\be_0$.

\noindent {\bf Remark:} Disregarding three special cases, all
conjugacy classes of $\Lpo$ are of this form.  This can be seen by
proceeding to the covering group $SL(2,\CC)$ of $\Lpo$ and making use
of the Jordan normal form of two--by--two matrices.

     One finds by explicit computation (most conveniently in the 
covering group) that the commutant of $\Lambda$ is
equal to the abelian group $\Lambda_1 {\cal L}_0 \Lambda_1^{-1}$.
Hence if $\be$ is such that $\be \cdot \be_0 = 0$ and if $\lambda$ is
the reflection about the edge of $W_{\be}$, one obtains for the
reflection $\lambda_1 \doteq \Lambda_1 \lambda \Lambda_1^{-1}$ about
the edge of $\Lambda_1 W_{\be}$ the equality
\begin{equation}
\lambda_1 \Lambda^\prime = \Lambda^{\prime \, -1} \lambda_1 \, \quad
\Lambda^\prime \in \Lambda_1 {\cal L}_0 \Lambda_1^{-1} \ .
\end{equation}
Hence $\blambda^\prime = \Lambda_1 {\cal L}_0 \Lambda_1^{-1}$ in this
case. Moreover, as $ {\cal L}_0$ is stable under taking square roots,
there is also for each $\Lambda^\prime \in \Lambda_1 {\cal L}_0
\Lambda_1^{-1}$ a square root 
$\Lambda^{\prime \, 1/2} \in \Lambda_1 {\cal L}_0 \Lambda_1^{-1}$.  
Hence for these special elements
$\Lambda$ we have complete control of their representation in terms of
products of reflections. We summarize these results in the following
lemma.
\begin{lemma} \label{ambiguity} Let 
$\Lambda \in \Lambda^\prime {\cal L}_0 \Lambda^{\prime \, -1}$,
$\Lambda^2 \neq 1$, and let $\lambda_1$
be the reflection about the edge of $\Lambda^\prime W_{\be}$,
where $\be$ is orthogonal to $\be_0$. Then 
$\lambda_2 = \lambda_1 \Lambda$
is a reflection and $\lambda_1 \lambda_2 = \Lambda$.
Moreover, any pair of reflections $\lambda_1^\prime, 
\lambda_2^\prime$ with product $\Lambda$ arises from 
$\lambda_1, \lambda_2$ by the adjoint action of some 
element of $ \Lambda^\prime {\cal L}_0 \Lambda^{\prime \, -1}$.
\end{lemma}
\subsection{Reflection maps}

     Let ${\cal J}$ be a topological group. 
We consider maps $\lambda \mapsto J(\lambda)$ from the set of
reflections $\Rs \subset \Lp$ into ${\cal J}$. There is no loss of
generality to assume that the subgroup generated by the set of elements 
$\{ J(\lambda) \mid \lambda \in \Rs \}$ is dense in ${\cal J}$.

\begin{definition} A map $J : \Rs \rightarrow \Js$ is a reflection 
map if for every $\lambda \in \Rs$ the element 
$J(\lambda) \in \Js$ is an involution and 
\begin{equation}
J(\lambda_1) J(\lambda_2) J(\lambda_1) = J(\lambda_1 \lambda_2 \lambda_1)
\ , \label{jcov}
\end{equation}
for all $\lambda_1, \lambda_2 \in \Rs$.
\end{definition}
We want to show that any continuous reflection map is the restriction of a
continuous homomorphism $V : \Lp \rightarrow {\cal J}$.

     For this to be true, it would be necessary to define for given 
$\Lambda \in \Lpo$ and corresponding pair of reflections 
$\lambda_1, \lambda_2$ with $\lambda_1 \lambda_2 = \Lambda$ the element
\begin{equation}
V(\Lambda) \doteq J(\lambda_1) J(\lambda_2) \ .
\end{equation}
Yet it is {\it a priori} not clear whether this element (a)
is independent of the choice of the pair of reflections into which
$\Lambda$ is decomposed, (b)
has the right continuity properties and (c) defines a homomorphism.
We shall start by making specific choices of reflections  
for special $\Lambda$ and establish, step by step, properties (a)--(c)
of $V$. In this discussion we make 
use of arguments and results in \cite{BDFS}, which we shall recall
here in somewhat modified form for the convenience of the reader.

     Let us consider first the action of $V$ on rotations and boosts, 
$R$, $B$. To this end we choose a vector $\be$ which is orthogonal to 
the axis of revolution of $R$, respectively the boost direction of $B$;
such vectors are called admissible in the following. 
Let $\lambda_\sbe$, $R \lambda_\sbe$ and
$B \lambda_\sbe$ be the reflections defined above. We then set
\begin{equation} \label{rotationsandboosts}
V_{\sbe}(R) \doteq J(R \lambda_\sbe) J(\lambda_\sbe) \ , 
\quad V_{\sbe}(B) \doteq J(B \lambda_\sbe) J(\lambda_\sbe) \ 
\end{equation}
and observe that $V_{\sbe}(R)^{-1} = J(\lambda_\sbe) J(R \lambda_\sbe)$
and $V_{\sbe}(B)^{-1} = J(\lambda_\sbe) J(B \lambda_\sbe)$.
Note that because of relation (\ref{jcov}), we also have
\begin{equation}
V_{\sbe}(R) J(\lambda) V_{\sbe}(R)^{-1} =  
J(R \lambda_\sbe) J(\lambda_\sbe) J(\lambda) J(\lambda_\sbe) J(R \lambda_\sbe) 
= J(R \lambda R^{-1}) \ , \label{rcov}
\end{equation}
for every $\lambda \in \Rs$, where we have used 
$\lambda_\sbe R \lambda_\sbe = R^{-1}$. Similarly, we also have
\begin{equation}
V_{\sbe}(B) J(\lambda) V_{\sbe}(B)^{-1} = J(B \lambda B^{-1}) \ .
\label{bcov}
\end{equation}
\begin{lemma} \label{independence}
The elements $V_{\sbe}(B)$, $V_{\sbe}(R)$ defined above
do not depend on the choice of the vector $\be$ within the
above-stated limitations. \label{inv}
\end{lemma}
\begin{proof} Consider first the case of boosts. If $B = 1$, there is
nothing to prove. So let $B \neq 1$, $\be$ be one of the admissible 
vectors for this boost, and let $B_1$ be any boost in the same direction 
as that of $B$. Note that $V_{\sbe}(B_1)J(\lambda_{\sbe}) = 
J(B_1 \lambda_{\sbe}) J(\lambda_{\sbe})^2 = J(B_1 \lambda_{\sbe}) =
J(\lambda_{\sbe})^2 J(B_1 \lambda_{\sbe}) =
J(\lambda_{\sbe}) V_{\sbe}(B_1)^{-1}$. Hence, for any $n \in \IN$ 
$$
V_{\sbe}(B_1)^{2n} J(\lambda_{\sbe}) = 
V_{\sbe}(B_1)^n J(\lambda_{\sbe}) V_{\sbe}(B_1)^{-n} = 
J(B_1{}^n \lambda_{\sbe} B_1{}^{-n}) = J(B_1{}^{2n} \lambda_{\sbe}) \ ,
$$
using (\ref{bcov}). Consequently, one has
$$
V_{\sbe}(B_1)^{2n} = V_{\sbe}(B_1)^{2n} J(\lambda_{\sbe})^2 = 
J(B_1{}^{2n} \lambda_{\sbe}) J(\lambda_{\sbe}) =  V_{\sbe}(B_1{}^{2n}) \ .
$$
Similarly, one sees that 
$$
V_{\sbe}(B_1)J(B_1 \lambda_{\sbe}) =
J(B_1 \lambda_{\sbe}) J(\lambda_{\sbe}) J(B_1 \lambda_{\sbe}) =
J(B_1 \lambda_{\sbe}) V_{\sbe}(B_1)^{-1}
$$ 
and therefore
\begin{equation*}
\begin{split}
V_{\sbe}(B_1)^{2n+1} 
& = V_{\sbe}(B_1)^{2n} J(B_1 \lambda_{\sbe}) J(\lambda_{\sbe}) 
 = V_{\sbe}(B_1)^{n} J(B_1 \lambda_{\sbe}) V_{\sbe}(B_1)^{-n} J(\lambda_{\sbe}) \\ 
& =  J(B_1^{2n+1} \lambda_{\sbe}) J(\lambda_{\sbe}) = V_{\sbe}(B_1^{2n+1})
\ .
\end{split}
\end{equation*}
Thus, one has $V_{\sbe}(B_1)^{n} = V_{\sbe}(B_1{}^n)$, for all $n \in \IN$.

     Now let $R_\phi$ be a rotation by $\phi$ about the axis established by the
direction of the boost $B$. Since $R_\phi$ and $B_1$ commute,
one obtains from relation (\ref{rcov}) 
\begin{equation*}
\begin{split} 
& V_{\sbe}(R_\phi) V_{\sbe}(B_1) V_{\sbe}(R_\phi)^{-1} 
   =   V_{\sbe}(R_\phi)J(B_1 \lambda_{\sbe}) 
     J(\lambda_{\sbe})V_{\sbe}(R_\phi)^{-1} \\
& =
J(B_1 R_\phi \lambda_{\sbe} R_\phi^{-1}) J(R_\phi \lambda_{\sbe} R_\phi^{-1}) 
=  J(B_1 \lambda_{ R_\phi \sbe}) J(\lambda_{ R_\phi \sbe}) 
= V_{R_\phi \sbe}(B_1) \ . 
\end{split}
\end{equation*} 
     On the other hand, according to (\ref{bcov}), the element
$V_{R_\phi \sbe}(B_1) V_{\sbe}(B_1)^{-1}$ must commute with $J(\lambda)$,
for every $\lambda \in \Rs$. Since $J(\Rs)$ generates $\Js$, this implies
that there exists some element $Z_{\phi}$ in the center of $\Js$ such that 
$$ V_{R_\phi \sbe}(B_1) =  Z_\phi V_{\sbe}(B_1) \ . $$
Setting $\phi = 2m \pi / n$, for $n \in \IN$ and $m \in \ZZ$, one sees 
from the preceding two relations that 
$$ 
V_{\sbe}(B_1) =  V_{R_{2m \pi / n}^{\, n} \sbe}(B_1)  =
 V_{\sbe}(R_{2m \pi / n})^{n} \, V_{\sbe}(B_1) V_{\sbe}(R_{2m \pi / n})^{-n} = 
Z_{2m \pi / n}^{\, n} V_{\sbe}(B_1) \ , 
$$
and consequently $Z_{2m \pi / n}^{\, n} = 1$. Hence,
$$ 
V_{R_{2m \pi / n} \sbe}(B_1^{\, n}) = V_{R_{2m \pi / n} \sbe}(B_1)^{\, n} 
=  Z_{2m \pi / n}^{\, n} V_{\sbe}(B_1)^{n} = V_{\sbe}(B_1)^{n} 
= V_{\sbe}(B_1^{\, n}) \ , 
$$
and setting $B_1 = B^{1/n}$ one obtains
$$ V_{R_{2m \pi / n} \sbe}(B) =  V_{\sbe}(B) \ , $$ 
for all $n \in \IN$, $m \in \ZZ$.  
By hypothesis, the reflection map is continuous, so the element 
$V_{\sbe}(R_\phi)$ depends continuously on $\phi$ for any admissible $\be$, 
and the same is thus also true of $V_{R_\phi \sbe}(B)$. It therefore 
follows from the preceding relation that 
$V_{R_\phi \sbe}(B) = V_{\sbe}(B)$ for any rotation 
$R_\phi$, proving the assertion for the case of the boosts.

     For the rotations $R$, one proceeds in exactly the same way. The 
role of $R_\phi$ is now to be played by the rotations about the axis 
of revolution fixed  by $R$.
\end{proof}

     In light of this result, we may omit the index $\be$ and set
\begin{equation}
V(B) \doteq V_{\sbe}(B) \ , \qquad V(R) \doteq V_{\sbe}(R) \ .
\end{equation}
\begin{lemma} The elements $V(B)$ and $V(R)$ depend continuously on
the boosts $B$ and rotations $R$, respectively. 
\end{lemma}
\begin{proof} Let $\{ B_n \}_{n \in \IN}$ be a sequence of boosts
converging to $B$. If $B \neq 1$ the distance between the unit disks
parameterizing the corresponding orthogonal admissible vectors converges to $0$.
In particular, there exists a sequence of unit vectors $\be_n$,
admissible for $B_n$, converging to the unit vector $\be$, admissible
for $B$. Hence, the sequence $\{ B_n \be_n \}$ converges to $B \be$.
By the assumed continuity of the reflection map, one concludes
that $V(B_n) = J(B_n \lambda_{\sbe_n})J(\lambda_{\sbe_n})$ converges 
to $J(B \lambda_\sbe)J(\lambda_\sbe) = V(B)$ 
as $n \rightarrow \infty$. 

     If, on the other hand, the sequence $\{ B_n \}$ converges to
$1$, the corresponding unit disks need not converge. Nonetheless,
due to the compactness of the unit ball in $\RR^3$, for any sequence
of unit vectors $\be_n \in \RR^3$ there exists a subsequence
$\{\be_{\sigma(n)}\}$ which converges to some unit vector $\be_\sigma$.
Since $\{ B_{\sigma(n)}\}$ converges to 1, the corresponding sequence
$\{ B_{\sigma(n)} \be_{\sigma(n)} \}$ converges to $\be_\sigma$. One 
therefore has
$$
V(B_{\sigma(n)}) = 
J(B_{\sigma(n)} \lambda_{\sbe_{\sigma(n)}})J(\lambda_{\sbe_{\sigma(n)}})
\rightarrow J(\lambda_{\sbe_\sigma}) J(\lambda_{\sbe_\sigma}) = 1 \ .
$$
Since the choice of sequence $\{ \be_n \}$ was arbitrary, the proof of 
the continuity of $V(B)$ with respect to the boosts $B$ is complete. The 
argument for the rotations is analogous after the boost direction is
replaced by the axis of revolution of the respective rotation.
\end{proof}
\begin{lemma} With the above definitions, one has the following.
\begin{itemize}
\item[(1)] $V(R) V(B) V(R)^{-1} = V(RBR^{-1})$ for all boosts 
$B$ and rotations $R$. 
\item[(2)]  $V(\cdot)$ defines a true representation of every continuous 
one-parameter subgroup of boosts or rotations.   
\end{itemize}
\end{lemma}
\begin{proof} Statement (1) follows from relation (\ref{rcov}) 
and Lemma \ref{inv}, which imply ($\be$ being admissible for both 
the rotation $R$ and the boost $B$)
\begin{equation*}
\begin{split}
V(R) V(B) V(R)^{-1} &  = 
V(R) J(B \lambda_\sbe) J(\lambda_\sbe) V(R)^{-1} \\ 
& = J(R B R^{-1} R \lambda_\sbe R^{-1}) J(R \lambda_\sbe R^{-1}) 
 =  J(R B R^{-1} \lambda_{R \sbe}) J(\lambda_{R \sbe})  \\ 
& =  V(RBR^{-1}) \ . 
\end{split}
\end{equation*}
The last equality follows from the fact that $RBR^{-1}$ is 
again a boost whose direction is orthogonal to 
$R \be $ .  

     Now let $G: \RR \rightarrow \Lpo$ be a continuous 
one-parameter group of 
boosts or rotations. As in the proof of Lemma \ref{inv}, one shows by an 
elementary computation on the basis of relation (\ref{jcov}) that 
$V(G(u))^n = V(G(u)^n)=V(G(nu))$. Consequently, one finds that, for 
$m_1,m_2,n \in \IN $,
\begin{equation*} 
\begin{split} 
V(G(m_1/n)) V(G(m_2/n)) & =  V(G(1/n))^{m_1} V(G(1/n))^{m_2} \\
& =  V(G(1/n))^{m_1+m_2} = V(G((m_1+m_2)/n)) \ . 
\end{split}
\end{equation*}
As $V(G(-u)) = V(G(u)^{-1}) = V(G(u))^{-1}$ by 
(\ref{rotationsandboosts}) and Lemma 2.4, this relation extends
to arbitrary $m_1,m_2 \in \ZZ$ and $n \in \IN $.
The stated assertion (2) thus follows once again from the continuity 
properties of $V(\, \cdot \,)$ established so far.  
\end{proof}

     Given any $\Lambda \in \Lpo$ we make use of its unique polar
decomposition $\Lambda = RB$ and choose a direction $\be$
which is orthogonal to both the axis of revolution of $R$
and the boost direction of $B$. We recall that the corresponding
reflection $\lambda_\sbe$ satisfies both $R \lambda_\sbe = \lambda_\sbe
R^{-1}$ and  $B \lambda_\sbe = \lambda_\sbe B^{-1}$. Hence 
$R \lambda_\sbe = RB \lambda_\sbe B$ and $\lambda_\sbe B$ are reflections, and 
we can define 
\begin{equation} \label{defi}
\begin{split}
V(\Lambda) \doteq & J( RB \lambda_\sbe B) J( \lambda_\sbe B)
= J(R \lambda_\sbe) J(\lambda_\sbe)^2  J( \lambda_\sbe B)
J(\lambda_\sbe)^2   \\ 
& = J(R \lambda_\sbe) J(\lambda_\sbe) J(B \lambda_\sbe)  J(\lambda_\sbe) = 
V(R) V(B) \ , 
\end{split}
\end{equation}
where we made use of the properties of reflection maps.
Since $R$, $B$ depend continuously on $\Lambda$, $V(\Lambda)$
is continuous in $\Lambda$ as well. Moreover, 
\begin{equation} \label{cova}
V(\Lambda) J(\lambda) V(\Lambda)^{-1} = J(\Lambda \lambda \Lambda^{-1})
\ ,
\end{equation}
for any $\lambda \in \Rs$ and $\Lambda \in \Lpo$.

     Let ${\cal L}_0 \subset \Lid$ be the abelian stability group of any given
wedge $W_{\sbe_0}$. It is generated by two one-parameter subgroups:
the rotations about the axis fixed by $\be_0$ and the boosts in the 
corresponding direction. It follows
from the properties of $V(\,\cdot\,)$ established so far that for any
$\Lambda = R B \in {\cal L}_0$ one has
\begin{equation}
V(R)V(B) = V(R)V(B)V(R)^{-1}V(R) = 
V(RBR^{-1})V(R) = V(B)V(R) \, .
\end{equation}
Hence for any 
$\Lambda_0 = R_0 B_0 \in {\cal L}_0$ one obtains
\begin{equation} \label{com}
V(\Lambda_0) V(\Lambda) V(\Lambda_0)^{-1} =  
V(R_0) V(B_0) V(R) V(B) V(B_0)^{-1} V(R_0)^{-1}
= V(\Lambda) \ . 
\end{equation}
This fact puts us into the position of being able to prove that for a 
large set of Lorentz transformations $\Lambda \in \Lpo$ the corresponding 
$V(\Lambda)$ do not depend on the choice of reflections in the decomposition 
of $\Lambda$. 
\begin{lemma} 
Let $\Lambda \in \Lambda^\prime {\cal L}_0 \Lambda^{\prime \, -1}$,
$\Lambda^2 \neq 1$, $\Lambda^\prime \in \Lid$. 
Then $V(\Lambda) = J(\lambda_1) J(\lambda_2)$ for any 
pair of reflections $\lambda_1$, $\lambda_2$ satisfying
$\lambda_1 \lambda_2 = \Lambda$.
\end{lemma}
\begin{proof} In view of relation (\ref{cova}) it suffices to
establish the statement for $\Lambda \in {\cal L}_0$, ${\Lambda}^2 \neq 1$. 
Let $\lambda_1, \lambda_2$ be reflections as in definition
(\ref{defi}) such that $\Lambda = \lambda_1 \lambda_2$ and 
$V(\Lambda) = J(\lambda_1) J(\lambda_2)$.
If $\lambda_3, \lambda_4$ are reflections 
such that $\lambda_3 \lambda_4 = \lambda_1 \lambda_2$, there is 
by Lemma~\ref{ambiguity} a $\Lambda_0 \in {\cal L}_0$ such that 
$\lambda_3 = \Lambda_0 \lambda_1 \Lambda_0^{-1}$, 
$\lambda_4 = \Lambda_0 \lambda_2 \Lambda_0^{-1}$. 
Hence by relations (\ref{cova}) and (\ref{com}) 
one obtains 
$$
J(\lambda_3) J(\lambda_4) = V(\Lambda_0) 
J(\lambda_1) J(\lambda_2) V(\Lambda_0)^{-1} = 
V(\Lambda_0) V(\Lambda) V(\Lambda_0)^{-1} 
= V(\Lambda) \ , 
$$
proving the statement.
\end{proof}

     This result will greatly simplify the computations which will show that
$V(\,\cdot\,)$ is a homomorphism. Let $R_1$, $R_2$ be arbitrary
rotations such that $(R_1 R_2)^2 \neq 1$ and let $\be$ be
orthogonal to the axes of revolution of  $R_1$, $R_2$. 
Taking into account the fact that any rotation is an element of the 
stability group ${\cal L}_0$ of some suitable wedge $W_{\sbe_0}$,
we obtain from the preceding lemma the equalities 
\begin{equation}
\begin{split}
V(R_1) V(R_2) &  =  J(R_1 \lambda_\sbe) J(\lambda_\sbe)
 J(R_2 \lambda_\sbe) J(\lambda_\sbe)  \\ & =
J(R_1 \lambda_\sbe) J(\lambda_\sbe R_2) = V(R_1 R_2) \, , 
\end{split}
\end{equation}
and this equation extends by continuity to arbitrary
pairs of rotations. Next, let $B_1$, $B_2$ be arbitrary boosts.
Then $B_1 B_2 = {B_1}^{1/2} ({B_1}^{1/2} B_2 {B_1}^{1/2})
{B_1}^{-1/2}$, where the expression in brackets is 
a positive matrix and hence a boost.
So $B_1 B_2$ belongs to the class of Lorentz transformations covered
by the preceding lemma. Choosing $\be$ orthogonal to the 
boost directions of $B_1$, $B_2$, we therefore have 
\begin{equation}
\begin{split}
V(B_1 B_2) & = J(B_1 \lambda_\sbe) J(\lambda_\sbe B_2) \\
& = J(B_1 \lambda_\sbe) J(\lambda_\sbe)^2 J(\lambda_\sbe B_2) 
J(\lambda_\sbe)^2  = V(B_1) V(B_2) \ .
\end{split}
\end{equation}
On the other hand, proceeding to the polar decomposition 
$B_1 B_2 = R B$ we get by definition $ V(B_1 B_2) = V(R) V(B)$ and hence 
\begin{equation}
V(B_1) V(B_2) = V(R) V(B) \ .
\end{equation}
Now let $\Lambda_1 = R_1 B_1$, $\Lambda_2 = R_2 B_2$ be arbitrary 
proper orthochronous Lorentz transformations. Introducing the boost
$B_{3} = R_2^{-1} B_1 R_2$ and making use of the polar
decomposition $B_{3} B_2 = RB$, we obtain from the  preceding 
results the chain of equalities
\begin{equation}
\begin{split}
V(\Lambda_1) V(\Lambda_2) =  
V(R_1) V(B_1) V(R_2) V(B_2) = V(R_1) V(R_2) V(B_{3}) V(B_2)
\\ = V(R_1 R_2) V(R) V(B) 
= V(R_1 R_2 R) V(B) = V(R_1 R_2 R B) = V(\Lambda_1 \Lambda_2) \ .
\end{split}
\end{equation}
Thus $V(\,\cdot\,)$ is a continuous homomorphism from $\Lpo$
into ${\cal J}$. It remains to extend  $V(\,\cdot\,)$ to the 
component of $\Lp$ which is disconnected from unity. To this end we fix 
a reflection $\lambda_0 \in \Lp$ corresponding to some wedge $W_{\sbe_0}$
and note that all elements in the disconnected
part can be represented uniquely in the form $\lambda_0 \Lambda$,
where $\Lambda \in \Lpo$. We set
\begin{equation}
V(\lambda_0 \Lambda) \doteq J(\lambda_0) V(\Lambda) \ .
\end{equation}
In view of the defining properties of reflection maps
and the definition of $V(\,\cdot\,)$, we get
\begin{equation} 
J(\lambda_0) V(\Lambda)  J(\lambda_0) = V(\lambda_0 \Lambda \lambda_0)
\ . 
\end{equation}
(Note that the sets of rotations and boosts
are mapped onto themselves by the adjoint action of $\lambda_0$,
and the set of distinguished wedges $W_\sbe$ is stable under the 
action of $\lambda_0$, as well.) Hence for any $\Lambda^\prime \in \Lpo$
\begin{equation}
\begin{split}
V(\Lambda^\prime) V(\lambda_0 \Lambda) & = J(\lambda_0)^2
V(\Lambda^\prime)  J(\lambda_0) V(\Lambda) =
J(\lambda_0) V(\lambda_0 \Lambda^\prime \lambda_0)V(\Lambda) \\
& = J(\lambda_0) V(\lambda_0 \Lambda^\prime \lambda_0 \Lambda) 
= V(\Lambda^\prime \lambda_0 \Lambda) \ , 
\end{split}
\end{equation}
and similarly $ V(\lambda_0 \Lambda)  V(\Lambda^\prime) 
=  V(\lambda_0 \Lambda \Lambda^\prime) $. Moreover, 
\begin{equation}
\begin{split}
V(\lambda_0 \Lambda) V(\lambda_0 \Lambda^\prime)
& =  J(\lambda_0) V(\Lambda) J(\lambda_0) V(\Lambda^\prime) \\
& = V(\lambda_0 \Lambda \lambda_0)  V(\Lambda^\prime) 
= V(\lambda_0 \Lambda \lambda_0 \Lambda^\prime) \ .
\end{split}
\end{equation}
Thus  $V(\,\cdot\,)$ is a continuous homomorphism from $\Lp$ into ${\cal J}$.

     As any reflection $\lambda$ can be represented in the form
$\lambda = \Lambda \lambda_0 \Lambda^{-1}$ for some $\Lambda \in \Lpo$,
it follows that 
\begin{equation}
J(\lambda) =  J(\lambda_0)^2 V(\Lambda) J(\lambda_0)  V(\Lambda)^{-1}
=  J(\lambda_0) V(\lambda_0 \Lambda \lambda_0 \Lambda^{-1}) =
V(\lambda) \ .
\end{equation}
Thus we finally see that  $J(\,\cdot\,)$  
is indeed the restriction of the continuous homomorphism  $V(\,\cdot\,)$  
to the set of reflections $\Rs$ and that $V(\Lambda)$ does not depend on the 
decomposition of $\Lambda$ into reflections for any $\Lambda \in \Lp$.
Since $\Rs$ generates $\Ls_+$, $V$ is the only
extension of the reflection map to a homomorphism from $\Ls_+$ into
$\Ts$. We summarize these results in the following proposition.
\begin{prop} \label{restriction} Let $J$ be a continuous reflection map 
from the set of reflections $\Rs \subset \Lp$ 
into an arbitrary topological group ${\cal J}$. Then $J$ is the 
restriction to $\Rs$ 
of a unique continuous homomorphism mapping  $\Lp$ into ${\cal J}$.
\end{prop}
\section{Continuity of modular reflection maps}

\setcounter{equation}{0}

     In view of the preceding proposition it is of interest to clarify 
the continuity properties of the modular reflection maps appearing in 
quantum field theory. In order to reveal the pertinent structures,
we discuss this problem in a setting which is slightly more general 
than that outlined in the introduction. 

     Let $W \mapsto \Rs(W)$ be any net of von Neumann algebras 
indexed by wedge regions, which satisfies the condition of 
wedge duality, $\Rs(W)^\prime = \Rs(W^\prime)$, and let 
$\Omega \in {\cal H}$ be any vector which is cyclic and separating
for all algebras $\Rs(W)$. We denote the modular conjugation
corresponding to the pair $\big(\Rs(W), \Omega\big)$ by $J_W$.
We shall show that the map 
$W \mapsto J_W$ from the family of wedges $\Ws$ into 
the group of (anti)unitary operators on $\Hs$ is continuous 
under quite general conditions. Making use of the fact that 
$\Pid$ acts transitively on $\Ws$, we identify $\Ws$, as 
a topological space, with the quotient space 
$\Pid / \Ps^{}_0 $, where $\Ps^{}_0 \subset \Pid$ 
is the invariance subgroup of any given wedge $W_0 \in \Ws$;
note that the topology does not depend on the choice
of $W_0$. On the group of (anti)unitary 
operators we use the strong--*--topology. 

     As we shall see, the desired result follows from the assumption that
the net $W \mapsto \Rs(W)$ is locally generated in the 
following specific sense: Let $\Cs$ be a family of closed
regions $C \subset \RR^4$ subject to the conditions: 
\begin{enumerate}
\item[(a)] Each $C \in \Cs$ can be approximated from the outside
by wedges $W \in \Ws$, \ie $C = \bigcap_{\, W \Supset \, C} W$.  
Here the inclusion relation $W \Supset C$ means that there is some open 
neighborhood of $W$ in $\Ws$ all of whose elements contain $C$.
\item[(b)] Each wedge $W \in \Ws$ can be approximated from the inside
by regions $C \in \Cs$, \ie $W = \bigcup_{\, C \Subset W} C$,
where $C \Subset W$ if $C$ is contained in all wedges 
in some neighborhood of $W$.  
\item[(c)] The family $\Cs$ is stable under the action of $\Pid$.
\end{enumerate}
We say in this case that $\Cs$ is a generating family of regions. 
A familiar example of such a generating family 
is the set of closed double cones in Minkowski space; another one
is the family of closed spacelike cones considered in \cite{BuFr} 
in the context of theories with topological charges. 

Given a generating family $\Cs$,
we define corresponding algebras $\Rs(C)$, $C \in \Cs$, setting
\begin{equation}
\Rs(C) \doteq \bigwedge_{W \Supset C} \Rs(W) \, .
\end{equation}
Clearly, $\Rs(C) \subset \Rs(W)$ whenever $C \Subset W$.

{\noindent \bf Definition:} The net $W \mapsto \Rs(W)$ is said to be 
locally generated if there is a generating family $\Cs$ of regions
such that $\Omega$ is cyclic for $ \Rs(C)$, $C \in \Cs$, and 
\begin{equation}
\Rs(W) = \bigvee_{C \Subset W} \Rs(C), \quad \ W \in \Ws \, .
\end{equation}

\noindent Note that the nets affiliated with quantum field 
theories satisfying the Wightman axioms are locally generated \cite{SWi}. 

We shall establish the continuity properties
of the modular conjugations by first showing that any
locally generated net satisfying wedge duality complies with the 
net continuity condition introduced in \cite{BDFS}, see below. 
Let $\{W_\delta\}_{\delta >0}$ be a family of wedges
converging to some wedge $W_0$ as $\delta$ converges to 0. 
We define corresponding von Neumann algebras  
\begin{equation} \label{lowerupper}
\underline{\Rs}_\varepsilon \doteq \bigwedge_{0  \leq \delta \leq \varepsilon}
\Rs(W_\delta) \, , \qquad 
\overline{\Rs}_\varepsilon \doteq \bigvee_{0 \leq \delta \leq \varepsilon}
\Rs(W_\delta) \, ,
\end{equation}
and note that, by construction,
\begin{equation} \label{zip}
\underline{\Rs}_\varepsilon \subset \Rs(W_\delta) \subset
\overline{\Rs}_\varepsilon \, ,
\end{equation}
for any $0 \leq \delta \leq \varepsilon$. Moreover, for any
$\varepsilon_1 \geq \varepsilon_2 \geq 0$, one has 
$\underline{\Rs}_{\varepsilon_1} \subset \underline{\Rs}_{\varepsilon_2}$ 
and 
$\overline{\Rs}_{\varepsilon_1} \supset \overline{\Rs}_{\varepsilon_2}$. 

Now let $C \in \Cs$ be such that $C \Subset W_0$. 
Bearing in mind the meaning of this 
inclusion relation, it is apparent that $C \Subset W_\delta$ for
all sufficiently small $\delta > 0$ and consequently  
$\Rs(C) \subset \underline{\Rs}_\varepsilon$ for sufficiently 
small $\varepsilon > 0$. Hence, 
$\Rs(C) \subset  \bigvee_{\varepsilon > 0} \underline{\Rs}_\varepsilon$.
As $ \bigvee_{\varepsilon > 0} \underline{\Rs}_\varepsilon$ is a 
von Neumann algebra and the net is locally generated, it follows that  
$\Rs(W_0) = \bigvee_{C \Subset W_0} \Rs(C)
\subset \bigvee_{\varepsilon > 0} \underline{\Rs}_\varepsilon$.
On the other hand, the inclusion (\ref{zip}) implies 
$ \bigvee_{\varepsilon > 0} \underline{\Rs}_\varepsilon \subset \Rs(W_0)$,
proving the equality
\begin{equation} \label{con1}
\bigvee_{\varepsilon > 0} \underline{\Rs}_\varepsilon = \Rs(W_0) \, .
\end{equation}
In a similar manner one shows that also   
\begin{equation} \label{con2}
\bigwedge_{\varepsilon > 0} \overline{\Rs}_\varepsilon = \Rs(W_0) \, ,
\end{equation}
since, by wedge duality, 
\begin{equation} 
\overline{\Rs}_\varepsilon^{\, \prime} = 
\bigwedge_{0 \leq \delta \leq \varepsilon} \Rs(W_\delta)^\prime =
\bigwedge_{0 \leq \delta \leq \varepsilon} \Rs(W_\delta^\prime) \, ,
\end{equation}
and it is easy to see that the family of wedges
$\{W_\delta^{\, \prime}\}^{}_{\delta >0}$ converges to $W_0^\prime$.
Applying the arguments in the preceding step, one obtains 
$\bigvee_{\varepsilon > 0} \overline{\Rs}_\varepsilon^{\, \prime} 
= \Rs(W_0^\prime) = \Rs(W_0)^\prime$. From this equality the assertion 
follows by taking commutants.

Relations (\ref{con1}) and (\ref{con2}) comprise the net continuity
condition used in \cite{BDFS} as a crucial technical assumption.
We have therefore shown that this condition is met by any net which 
is locally generated and satisfies wedge duality. This fact puts
us into the position to apply Proposition 4.6 in \cite{BDFS},  
giving the following unexpected result.\footnote{Although
the CGMA is mentioned in the statement of \cite[Prop.\ 4.6]{BDFS}, 
only wedge duality and the cyclicity properties of $\Omega$ 
given above enter into its proof.}
\begin{prop} \label{netcontinuity}
Let $W \mapsto \Rs(W)$ be a locally generated net which
satisfies the condition of wedge duality. Then the map $W \mapsto J_W$
from the wedges $W \in \Ws$ to the modular conjugations
$J_W$ corresponding to $\big(\Rs(W),\Omega\big)$ is continuous.
\end{prop}
The conditions underlying this result are natural 
from a physical point of view and obtain in many models. We 
therefore believe that the continuity of the modular 
conjugations is a generic feature in local quantum physics.
As a matter of fact, such continuity properties can be 
established for locally generated nets on many space--times, 
provided they satisfy an analogue of the wedge duality condition. 

It is necessary to prove a version of Proposition \ref{netcontinuity} 
formulated in more group theoretical terms. Because of wedge duality, 
the modular conjugations $J_W$, $J_{W^\prime}$ corresponding to the pairs  
$(\Rs(W), \Omega)$ and $(\Rs(W^\prime), \Omega)$, respectively, 
coincide. We may therefore return to the notation 
used in the introduction, \ie put 
$J(\lambda) \doteq J_W = J_{W^\prime}$, where 
$\lambda \in \Ps_+$ is the reflection 
about the common edge $E$ of the wedges $W$ and $W^\prime$. 
This notation is consistent, since every two--dimensional
spacelike plane $E$ determines uniquely a corresponding pair of 
wedges $W$, $W^\prime$ having $E$ as their edge. 

As we shall see, Proposition \ref{netcontinuity} implies that the map 
$\lambda \mapsto J(\lambda)$ from the family of reflections 
$\lambda \in \Ps_+$ into the group of (anti)unitary 
operators is continuous. In order to prove this fact, we
have to show that for any given family of 
reflections $\{\lambda_\delta\}_{\delta > 0}$ 
converging to some reflection $\lambda_0$ there 
exists a corresponding convergent family of wedges whose elements 
$W_\delta$ have edges $E_\delta$ which are pointwise fixed 
under the action of $\lambda_\delta$, ${\delta > 0}$. 

The edges $E_\delta$ can be specified easily: Let 
$E_0$ be the two--dimensional spacelike plane which is 
pointwise fixed under the action of $\lambda_0$. Introducing
on $\RR^4$ the maps 
$x \mapsto \frac{1}{2} (1 + \lambda_\delta \lambda_0) x$
one finds that the planes 
$E_\delta =  \frac{1}{2} (1 + \lambda_\delta \lambda_0) E_0$ 
are pointwise fixed under the action of the reflections 
$\lambda_\delta$, $\delta > 0$. Since 
$\lambda_\delta$  converges to $\lambda_0$ it follows
that $\lambda_\delta \lambda_0$  converges to the 
unit element of $\Ps_+$, so the above maps  
converge to the identity, uniformly on compact subsets of
$\RR^4$. Thus, for sufficiently small $\delta > 0$,
each $E_\delta$ is a two--dimensional spacelike
plane and therefore constitutes the edge of a pair 
of wedges. Moreover, since the points on the edges $E_\delta$ 
converge to points on $E_0$ in the limit of small $\delta$, 
it is straightforward to exhibit 
Poincar\'e transformations $\upsilon_\delta \in \Pid$ 
such that $\upsilon_\delta E_0 = E_\delta$, $\delta > 0$,
and $\{\upsilon_\delta \}_{\delta > 0}$ converges to the
identity. Picking one of the two wedges with edge $E_0$,
say $W_0$, we take as our family of wedges 
$\{W_\delta \doteq \upsilon_\delta W_0 \}_{\delta > 0}$ 
and note that it has all of the desired properties.
Since $J(\lambda_\delta) = J_{W_\delta}$,
we are now in a position to apply Proposition \ref{netcontinuity},
entailing the following result.
\begin{cor} \label{cor3.2}
Let $W \mapsto \Rs(W)$ be a locally generated net 
which satisfies the con\-dition of wedge duality. Then the map 
$\lambda \mapsto J(\lambda)$ from the reflections 
$\lambda \in \Ps_+$ to the modular conjugations 
$J(\lambda)$ corresponding to $\big(\Rs(W),\Omega\big)$,
hence also to $\big(\Rs(W^\prime),\Omega\big)$, is 
continuous. Here the wedges $W$, $W^\prime$ are fixed 
by the condition $\lambda W = W^\prime$.
\end{cor}
We conclude this section with a technical result pertaining to nets
which satisfy the condition of geometric modular action, CGMA, and the
modular stability condition, CMS; see the next section for their
definitions.  The latter condition says that the elements of the
modular groups $\Delta_W^{it}$, $t \in \RR$, corresponding to the
pairs $\big(\Rs(W),\Omega\big)$ are contained in the group generated
by all finite products of the modular conjugations $J^{}_W$, 
$W \in \Ws$. Under these circumstances one can relax the condition that the
nets are locally generated without changing the conclusions of the
preceding discussion. In fact, one has the following result.
\begin{lemma} \label{lemma3.3}
Let $W \mapsto \Rs(W)$ be a net 
satisfying the CGMA and the CMS. If there is a generating
family of regions $\Cs$ such that $\Omega$ is cyclic for $\Rs(C)$, 
given any $C \in \Cs$, then the net is locally generated.
\end{lemma}
\begin{proof} As was shown in \cite{BDFS}, the  CGMA entails
relation (\ref{covar}), hence the CMS implies that, 
for any given $W \in \Ws$,   
$$ 
\Delta_W^{it} \Rs(W_0)  \Delta_W^{-it} =  \Rs(\upsilon_W(t) W_0) \, ,
   \quad W_0 \in \Ws \, ,
$$
where $\upsilon_W (t) \in \Ps_+$ for $t \in \RR$; as a 
matter of fact, since
$\Delta_W^{it} = ( \Delta_W^{it/2} )_{}^2$, $\upsilon_W(t)$
is the square of an element of $\Ps_+$ and thus lies 
in the identity component of this group. Moreover, since the modular unitaries 
$\Delta_W^{it}$ induce automorphisms of $\Rs(W)$ and since, by the CGMA, 
the map $W \mapsto \Rs(W)$ is a bijection, one finds that $\upsilon_W (t)$
must be an element of the stability group of $W$, $t \in \RR$.

Now let $\Cs$ be a generating family of regions as hypothesized, 
and let $C \in \Cs$ be such that $C \Subset W$
for a given $W \in \Ws$. Bearing in mind the definition of the 
algebras $\Rs(C)$ and the stability of the family $\Cs$ under 
Poincar\'e transformations, one obtains
$\Delta_W^{it} \Rs(C) \Delta_W^{-it} = \Rs(\upsilon_W (t) C)$
and $\upsilon_W (t) C \in \Cs$. Since $\upsilon_W (t)$
is an element of the stability group of $W$, one also finds, after a moment's
reflection, that $\upsilon_W (t) C \Subset W$. It follows that 
$\Delta_W^{it} \big( \bigvee_{C \Subset W}  \Rs(C) \big) \Delta_W^{-it} =  
\bigvee_{C \Subset W}  \Rs(C)  \subset \Rs(W) $, $t \in \RR$.
But $\Omega$ is cyclic for the algebras $\Rs(C)$,
hence $\bigvee_{C \Subset W}  \Rs(C) = \Rs(W)$, by a well known
result of Takesaki.
\end{proof}

Analogous results can be established for other 
space--times, a prominent example being de--Sitter space.

\section{Modular action and Poincar\'e covariance}

\setcounter{equation}{0}

Making use of the results obtained in the preceding two sections, we
are now able to improve considerably on the analysis of theories
complying with the CGMA carried out in \cite{BFS,BDFS}.  For the
convenience of the reader, we recall this condition in a form
appropriate for the present discussion as well as some of the
important consequences established in the cited articles.

A state vector $\Omega$ and a net $W \mapsto \Rs(W)$ from the family of 
wedge regions $\Ws$ in Minkowski space to von Neumann algebras are 
said to comply with the CGMA if the following conditions are satisfied.

\begin{enumerate}
\item[(a)] $W \mapsto \Rs(W)$ is an order-preserving bijection.
\item[(b)] If $W_1 \cap W_2 \neq \emptyset$, then $\Omega$ is cyclic
and separating for $\Rs(W_1) \cap \Rs(W_2)$. Conversely, if
$\Omega$ is cyclic and separating for $\Rs(W_1) \cap \Rs(W_2)$,
then  $\overline{W_1} \cap \overline{W_2} \neq \emptyset$, where
the bar denotes closure. 
\item[(c)] For each $W \in \Ws$, the adjoint action of the modular
conjugation $J_W$ correspon\-ding to the pair $(\Rs(W),\Omega)$ leaves
the set $\wrnet$ invariant.
\item[(d)] The group of (anti)automorphisms generated by the 
adjoint action of the modular conjugations  $J_W$, $W \in \Ws$,
acts transitively on  $\wrnet$.
\end{enumerate}

The core of this condition is part (c), which says that
the modular conjugations generate a subgroup of the symmetric (permutation) 
group on the set $\wrnet$. So, in view of the correspondence 
between algebras and wedge regions, it is appropriate to say that
these conjugations act geometrically. No {\it a priori} assumptions
are made about the specific form of this action and the nature
of the resulting group. This general formulation of the condition 
is also appropriate if one thinks of applications 
to nets labelled by other families of regions appearing, for example, in 
theories on curved space--times.

As was shown in \cite{BDFS}, cf.\ also \cite{Bor3}, the CGMA implies
that the net satisfies wedge duality. So one may consistently label
these conjugations by the reflections $\lambda \in \Ps_+$ about the edges 
of the wedges $W$, respectively $W^\prime$, setting 
$J(\lambda) \doteq J_W = J_{W^\prime}$. Moreover, 
the modular conjugations act on the net covariantly in the 
sense of relation (\ref{covar}) and satisfy the fundamental 
relation (\ref{prop}), see \cite{BDFS}.

We make use now of the additional assumption that the net is locally 
generated. Then, by Corollary \ref{cor3.2},   
$\lambda \mapsto J(\lambda)$ is a continuous map from the 
set of reflections $\lambda \in \Ps_+$, equipped with the topology
induced by $\Ps_+$, to the group of (anti)unitary operators, equipped
with the strong--*--topology. 
Using this information, we want to construct a continuous representation 
of the semidirect product $\Ls_+ \ltimes \RR^4 = \Ps_+$ whose elements are
denoted, as usual, by $(\Lambda,x)$. 

Restricting attention first to the subgroup $\Ls_+$, we can apply
Proposition \ref{restriction} and extend the reflection maps
$\lambda \mapsto J(\lambda,0)$, $\lambda \in \Ls_+$, 
to a continuous (anti)uni\-tary representation $U$ of the proper
Lorentz group $ \Ls_+$, $(\Lambda,0) \mapsto U(\Lambda,0)$.
Turning to the subgroup $\RR^4$ of translations, 
let $\lambda \in \Ls_+$ be any reflection and let 
$x \in \RR^4$ be such that $\lambda x = - x$. The 
set of such $x$ constitutes a two--dimensional 
subgroup of $\RR^4$ containing timelike translations.
Moreover, $(\lambda,x) = (1, x/2) (\lambda,0) (1, -x/2)$,
hence $(\lambda,x)$ is again a reflection. We put
\begin{equation} \label{transl}
U_\lambda (x) \doteq J(\lambda,x) J(\lambda,0) \, , \quad \lambda x = - x \, ,
\end{equation}
and note that it has been shown in \cite[Sec.\ 4.3]{BDFS} that $U_\lambda$
defines a continuous unitary representation of the two--dimensional
subgroup of translations $x$ satisfying the stated
condition.\footnote{This 
result was established in \cite{BDFS} only for reflections about
edges of certain specific wedges; the general statement follows from
the special case by an application of relation (\ref{prop}).}  As a
matter of fact, the continuity of this representation follows directly
from the CGMA without any further assumptions \cite{BFS}.  We want to
show next that these representations of subgroups of translations can
be combined with the above representation of the Lorentz group to
yield a true representation of the proper Poincar\'e group.

Let $x \in \RR^4$ be any timelike vector; so its stability subgroup 
in $\Ls_+$ is conjugate to the entire group of rotations. Hence,
applying the arguments in the proof of Lemma \ref{independence} to 
the present situation, one finds that for all elements $S \in \Ls_+$ of 
the stability group of $x$ one has 
$U_{S \lambda S^{-1}} (x) =  U_\lambda (x)$. We may 
therefore omit the dependence of these operators 
on the reflections $\lambda$ and set, for given timelike $x$,
\begin{equation}
U(1,x) \doteq U_{\lambda}(x) \, ,  
\end{equation}
for any reflection $\lambda$ such that $\lambda x = - x$. Now if $x$, $y$ are 
positive timelike, their sum is positive timelike, too, and there is 
a reflection $\lambda \in \Ls_+$ such that $\lambda x = -x$, 
$\lambda y = - y$. Hence we can compute
\begin{equation} \label{compute}
U(1,x) \, U(1,y) = U_\lambda(x) U_\lambda(y) = U_\lambda(x + y) = U(1,x + y) \, ,
\end{equation}
where, in the second equality, we made use of the fact established above
that $U_\lambda$ is a representation of the respective two--dimensional 
subgroup. We also note that, for timelike $x$ and admissible $\lambda$, 
\begin{equation} \label{inverse}
\begin{split}
U(1,x)^{-1} & =  J(\lambda,0) \, J(\lambda,x) =
J(\lambda,0)  \, J(\lambda,x) \, J(\lambda,0)^2 \\ 
& =   J(\lambda,-x) \, J(\lambda,0) = U(1,-x) \, ,
\end{split}
\end{equation}
where we made use of relation (\ref{prop}).
The preceding two relations allow us to extend the 
unitary operators $U(1,z)$ to arbitrary translations 
$z \in \RR^4$. Indeed, any $z$ can be decomposed into $z = x - y$, where 
$x$, $y$ are positive timelike, and, if 
$z = x^\prime - y^\prime$ is another such decomposition, 
one has $x + y^\prime = y + x^\prime$. Hence
$U(1,x)U(1,y^\prime) = U(1,x + y^\prime) = U(1,y + x^\prime) 
= U(1,y) U(1,x^\prime)$.
All operators appearing in this equality commute with each other
according to relation (\ref{compute}). Thus, making use of relation
(\ref{inverse}), one gets $U(1,x)U(1,-y) = U(1,x^\prime) U(1,-y^\prime)$.
One can therefore consistently define for $z \in \RR^4$
\begin{equation} \label{extend}
U(1,z) \doteq U(1,x) U(1,-y) \, , \quad x,y \in V_+ \, , x - y = z \, . 
\end{equation}
Based on the equalities established thus far, one can show
that $U$ defines a continuous unitary representation of the 
subgroup $\RR^4$ of translations. We omit the straightforward
proof of this fact.

Since timelike vectors $x$ are mapped by the elements $\Lambda \in \Ls_+$ 
to timelike vectors, one can also compute the adjoint action of the 
unitary operators $U(\Lambda,0)$ on $U(1,x)$. Fixing $x$ and making use 
of relations (\ref{transl}) and (\ref{prop}), one gets for any 
admissible reflection $\lambda$ the equation
\begin{equation} 
\begin{split}
U(\Lambda,0) \, U(1,x) \, U(\Lambda,0)^{-1} & = 
U(\Lambda,0) \, J(\lambda,x) \, J(\lambda,0) \, U(\Lambda,0)^{-1} \\
& =  J(\Lambda \lambda \Lambda^{-1}, \Lambda x) \, J(\Lambda \lambda \Lambda^{-1},0)
= U(1, \Lambda x) \, .
\end{split}
\end{equation}
This equation can be extended to arbitrary $x \in \RR^4$ 
by means of relation (\ref{extend}). Hence, setting
\begin{equation}
U(\Lambda,x) \doteq U(1,x) \, U(\Lambda,0) \, , \qquad (\Lambda,x) \in \Ps_+ \, ,
\end{equation}
we arrive at a continuous unitary representation of $\Ps_+$. 
Since for any reflection $\lambda \in \Ls_+$ and $x \in \RR^4$ 
such that $\lambda x = -x $ we have 
\begin{equation}
\begin{split}
U(\lambda,x) & = U(1,x) U(\lambda,0) =  U(1,x/2) U(\lambda,0)  U(1,x/2)^{-1} \\
& =   U(1,x/2) J(\lambda,0)  U(1,x/2)^{-1} =  J(\lambda,x) \, ,
\end{split}
\end{equation}
we also see that the representation $U$ extends the reflection
map $J$.

As a consequence of relation (\ref{covar}) and the fact that 
the operators $U(\Lambda,x)$ are certain specific products of modular 
conjugations, these operators act covariantly on 
the net. Moreover, because of the invariance of $\Omega$ under the 
action of the modular conjugations, $\Omega$ is invariant under
the action of $U(\Lambda,x)$, $(\Lambda,x) \in \Ps_+$. We summarize 
these results in the following theorem. 
\begin{theorem} \label{big}
Let $W \mapsto \Rs(W)$ be a locally generated net and $\Omega$
a state vector complying with the CGMA, \ie conditions 
(a) to (d). Then the net satisfies 
wedge duality and there is a
continuous (anti)unitary representation $U$ of $\Ps_+$ which
leaves $\Omega$ invariant and acts covariantly on the net.
Moreover, for any given wedge $W$ and reflection
$\lambda$ about its edge, $U(\lambda)$ is the modular 
conjugation corresponding to the pair $(\Rs(W), \Omega)$.
\end{theorem}
     Although $\Omega$ is invariant under the action of 
$U$ and as such clearly is a distinguished state, the CGMA does not imply 
that it is necessarily a ground state. In fact, there exist examples
conforming with the hypothesis of Theorem \ref{big} for 
which the joint spectrum $\mbox{sp}\, U$ of the generators of the  
subgroup of translations is all of $\RR^4$, cf. \cite[Sec.\ 5.3]{BDFS}. 
So, for the characterization of ground states on Minkowski space, 
where $\mbox{sp} \, U$ is contained in a light cone, 
one has to supplement the CGMA by additional constraints. 
A conceptually simple and quite general requirement
is the modular stability condition CMS, proposed in \cite{BDFS}. 
We recall this condition here as the 
last item in our list of constraints characterizing 
Poincar\'e invariant ground states describing the vacuum.
\begin{enumerate}
\item[(e)] For any $W \in \Ws$, the elements $\Delta_W^{it}$, $t \in \RR$, 
of the modular group corresponding to $(\Rs(W), \Omega)$ are contained 
in the group generated by all finite products of the modular
involutions $\{ J_W\}_{W \in \Ws}$.
\end{enumerate}
We refer the interested reader to \cite{BFS,BDFS} for a discussion of
the background of this condition and a brief account of other
interesting approaches towards an algebraic characterization of ground
states. Within the present context, the implications of condition (e)
are twofold. On the one hand, we can apply Lemma \ref{lemma3.3} and
replace in Theorem \ref{big} the assumption that the net is
locally generated by the weaker requirement that there is some
generating family $\Cs$ of regions such that $\Omega$ is cyclic for
the algebras $\Rs(C)$, $C \in \Cs$. On the other hand, we can employ
the results in \cite{BFS,BDFS}, which imply that the representation
$U$ has the desired spectral properties.
\begin{theorem}
Let $W \mapsto \Rs(W)$ be a net and $\Omega$
a state vector satisfying the CGMA and CMS, \ie conditions
(a) to (e), and let $\Cs$ be 
some generating family of regions such that $\Omega$ is cyclic
for the algebras $\Rs(C)$, $C \in \Cs$. Then the net
satisfies wedge duality and there is a 
representation $U$ of $\Ps_+$ with properties described
in the preceding theorem such that 
$\mbox{\rm sp}\, U  \subset \overline{V}_{\! +}$ or 
$\mbox{\rm sp}\, U \subset -\,\overline{V}_{\! +}$,
where $ \overline{V}_{\! +}$ denotes the closed forward
lightcone.
\end{theorem}
   The fact that both the forward and the backward 
lightcone appear as possible supports of the spectrum of the 
generators of the translations can be understood easily: Neither the CGMA
nor the CMS contains any input about the arrow of time.
By choosing proper coordinates, one may therefore
assume without loss of generality that $\mbox{\rm sp}\, U \subset
\overline{V}_{\! +}$.  With this convention, $U$ is then the only
continuous unitary representation of the spacetime translations which
acts covariantly on the given net and leaves $\Omega$ invariant 
\cite[Prop.\ 5.2]{BDFS}, cf.\ also \cite[Prop.\ 2.4]{BS}.
 
   We have thus attained our goals, the characterization of vacuum
states in Minkowski space and the construction of continuous unitary
representations of the isometry group of this space, using conditions
which are expressed solely in terms of the algebraically determined
modular objects. All technical assumptions about continuity properties
of the net have been eliminated from this analysis. Instead, they 
follow in a natural manner from the net structure. And Proposition
\ref{restriction} has made clear that once this continuity has been assured,
the modular reflection map determines the representation of the
isometry group in a canonical and unique manner. 
   
   There are already strong indications 
from studies of nets on de Sitter \cite{BDFS,F2} and anti-de Sitter 
space--times \cite{BFS1},  
and also more general Robertson--Walker space--times \cite{BMS1,BMS2}
that this strategy is applicable to a large class of space--times 
of physical interest. This is true in spite of the 
fact that there is no translation subgroup in the isometry group 
of these spaces, and thus the standard definition of vacuum state 
is inapplicable. We therefore believe that the analysis of the 
modular data in quantum field theories on curved space--time
deserves further attention.

\bigskip

\noindent {\bf Acknowledgements}:  
DB wishes to thank the Institute for Fundamental Theory of the University
of Florida and SJS wishes to thank the Institute
for Theoretical Physics of the University of G\"ottingen 
for hospitality and financial support which facilitated this research.
This work was supported in part by a research grant of 
Deutsche Forschungsgemeinschaft (DFG)

\end{document}